\documentclass[aps,prf,preprint,groupedaddress]{revtex4-2}

\usepackage{amsmath} 
\usepackage{graphicx}
\usepackage[]{siunitx}
\DeclareSIUnit\year{yr}
\DeclareSIUnit\wt{wt}
\DeclareSIUnit\gforce{\text{$g$}}

\usepackage{xcolor}

\newcommand{\vl}[2]{\textcolor{blue}}

\begin{document}


\title{On the concentration distribution in turbulent thermals}


\author{Ludovic Huguet}
\affiliation{ISTerre, Université Grenoble Alpes, Université Savoie Mont Blanc, CNRS, IRD, Université Gustave Eiffel, 38000 Grenoble, France}
\affiliation{School of Earth and Environment, University of Leeds, Leeds LS2 9JT, UK}

 \email{l.g.huguet@leeds.ac.uk}
\author{Victor Lherm}
\affiliation{Université Paris-Saclay, CNRS, FAST, 91405, Orsay, France}
\author{Renaud Deguen}
\affiliation{ISTerre, Université Grenoble Alpes, Université Savoie Mont Blanc, CNRS, IRD, Université Gustave Eiffel, 38000 Grenoble, France}
\author{Joris Heyman}
\affiliation{Univ. Rennes, CNRS, Géosciences Rennes (UMR6118), 35042 Rennes, France}
\author{Tanguy Le Borgne}
\affiliation{Univ. Rennes, CNRS, Géosciences Rennes (UMR6118), 35042 Rennes, France}



\date{\today}

\begin{abstract}
Turbulent thermals emerge in a wide variety of geophysical and industrial flows, such as atmospheric cumulus convection and pollutant dispersal in oceans and lakes. When a buoyant fluid mass rises, or sinks, heat and mass transfer occur through the engulfment of fresh surrounding fluid into the thermal --- a process that spans multiple scales from macroscopic entrainment of ambient fluid to microscopic diffusive processes. Turbulent thermals are typically investigated through their integral properties (radius, depth, and entrainment rate). However, mixing processes depend on the internal distribution of concentration or temperature inside a thermal, which remains poorly constrained. Here, we use laboratory fluid dynamics experiments and direct numerical simulations to investigate the mixing of a passive scalar in turbulent thermals with large Reynolds numbers. We track the evolution of the concentration field by computing its moments and the probability density function. The concentration distribution exhibits self-similarity over time, except at high concentrations, possibly because of the presence of undiluted cores. These distributions are well approximated by an exponential probability density function. Although diffusion has a strong effect on the spatial structure of the concentration field, we observe no significant effect of diffusivity on the concentration distributions in the investigated range of Péclet numbers.
\end{abstract}


\maketitle


\section{Introduction}
\label{sec:introduction}

A finite volume of buoyant fluid evolving under the influence of gravity in an otherwise quiescent environment forms what is called a 	\textit{turbulent thermal} when the flow is turbulent. The term 	\textit{thermal} originates from the terminology used by glider pilots, referring to a volume of warm air rising through the atmosphere \citep{s1957,w1958}. Turbulent thermals are fundamental elements of atmospheric convection and have been extensively studied by both the fluid dynamics and atmospheric science communities \citep{b1954,blc2005,yano2014basic}. 
Yet the interest in turbulent thermals also extends to other contexts. 
Flows sharing strong similarities with atmospheric thermals can also develop in liquid environments, without necessarily being driven by temperature-induced density variations. The nature of the buoyancy forces — whether due to temperature variations, compositional differences, or the presence of multiple phases with differing densities — appears to be of secondary importance for the overall behavior of the thermal \citep{b1954,t1979,btb2003,ldo2014}. 

Field observations and laboratory experiments show that turbulent thermals expand and decelerate as they rise, as a result of the gradual entrainment of ambient fluid \citep{b1954,s1957,w1959}. This is consistent with dimensional analysis \citep{b1954}, which predicts that far from the source the thermal radius $r$, velocity $w$, and reduced gravitational acceleration (or local buoyancy) $g'$ evolve with distance $z$ and time $t$ as
\begin{align}
r &\sim z \sim \mathcal{B}_0^{1/4} t^{1/2}, \label{eq:scaling_r} \\
w &\sim \mathcal{B}_0^{1/2} z^{-1} \sim \mathcal{B}_0^{1/4} t^{-1/2}, \label{eq:scaling_w} \\
g' = \frac{\Delta\rho}{\rho_a}g &\sim \mathcal{B}_0 z^{-3} \sim \mathcal{B}_0^{1/4} t^{-3/2} .\label{eq:scaling_g}
\end{align}
where $\rho_a$ is the density of the ambient fluid, $\Delta\rho$ is the difference between the average density of the thermal and the ambient at time $t$,  and $g$ is the acceleration of gravity.
$\mathcal{B}_0$ is the total buoyancy of the thermal, defined as
\begin{equation}
    \mathcal{B}_0=\frac{\Delta \rho_0}{\rho_a} g V_0,\, \text{ (m$^{4}\,$s$^{-2}$)}\label{eq:InitialBuoyancy}
\end{equation}
where $\Delta\rho_0$ is the value of $\Delta\rho$ at release, and $V_0$ the initial volume of the thermal. Equation \ref{eq:scaling_r} predicts that the size of the thermal increases in proportion to the distance traveled by the thermal, while equations \ref{eq:scaling_w} and \ref{eq:scaling_g} predict that the velocity and buoyancy decrease with distance and time, which results from the gradual dilution of the momentum and buoyancy of the thermal implied by entrainment of ambient fluid.

It is instructive to analyze the implications of equations \ref{eq:scaling_r}-\ref{eq:scaling_g} for the evolution with $z$ of the different terms of the momentum conservation equation. Assuming that velocity fluctuations vary over a length scale $\sim r$, a scale analysis of the different terms of the momentum conservation equation shows that inertia scales as $\mathbf{v}\cdot \mathbf{\nabla}  \mathbf{v} \sim  \mathcal{B}_0 z^{-3}$, buoyancy $g' \sim  \mathcal{B}_0 z^{-3}$, and viscous forces $\nu \mathbf{\nabla}^{2}  \mathbf{v} \sim \nu \mathcal{B}_0^{1/2} z^{-3}$ ($\nu$ denotes kinematic viscosity).
The ratio of inertia to viscous forces, which defines an instantaneous Reynolds number $Re$, is constant in time \citep{s1957,j1992,mnc1994} and given by
\begin{equation}
\frac{\text{inertia}}{\text{viscous force}} \sim 
Re = \frac{\mathcal{B}_0^{1/2}}{\nu}.
\label{eq:re}
\end{equation}
If the buoyancy originates from a concentration field $c$ with diffusivity $D$, one can infer that the scaling laws \ref{eq:scaling_r}-\ref{eq:scaling_g} (with $g'\propto c$) imply that the advection and diffusion terms $\mathbf{v}\cdot \mathbf{\nabla} c$ and $D \nabla^2 c$
of the transport equation for $c$ are both $\propto z^{-5}$ when estimated for concentration variations over a length scale $\sim r$. 
Their ratio, which defines an instantaneous Péclet number $Pe$, is constant and given by
\begin{equation}
\frac{\text{inertia}}{\text{diffusion}} \sim Pe = \frac{\mathcal{B}_0^{1/2}}{D}.\label{eq:pe}
\end{equation}
The same conclusion is reached if buoyancy originates from temperature rather than compositional variations. Note that the scaling given in equations \ref{eq:re} and \ref{eq:pe} for the Reynolds and Péclet numbers can also be obtained from their usual definitions, $Re=w r/\nu$ and $Pe=w r /D$, using the scaling laws \ref{eq:scaling_r} and \ref{eq:scaling_w} for $r$ and $w$.

The fact that the instantaneous Reynolds and Péclet numbers are constant suggests that turbulent thermals can evolve in a self-similar way when far from the source, in the sense that velocity and concentration or temperature would each have similar distributions at different times or distances from the source.
This is consistent with early experimental determinations of the mean velocity field \citep{w1959}, and with experimental observations showing that the shape of thermals is often self-preserved \citep{s1957}.

However, some experimental observations argue against self-similarity. Bond \& Johari (2010) \cite{bj2010} found that the vorticity within turbulent thermals tends to be localized in thin vortex cores, the width of which increases more slowly than the thermal radius.
This was confirmed by Zhao et al.~(2013) \cite[][]{zlla2013}, who also found that the maximum concentration within the thermal decreases at a much slower rate than predicted for the average concentration according to Eq.~(\ref{eq:scaling_g}). Note that the scaling laws (Eq.(\ref{eq:scaling_r}-\ref{eq:scaling_g})), which imply constant $Re$ and $Pe$, assume that the total buoyancy of the thermal is conserved. 
Experiments and numerical simulations show that detrainment (\textit{i.e.}, loss of mass and buoyancy) is small \citep{j1992,lj2019,m2025} but not inexistent (an order of magnitude smaller than entrainment according to \cite{lj2019}).
This implies that the instantaneous $Re$ and $Pe$ decrease with time, albeit slightly, which does not point toward asymptotic self-similarity.

The question of self-similarity of the concentration is also related to the dynamics of mixing and homogenization of thermal or compositional heterogeneities, which sets the distribution of temperature or concentration within the thermal. 
The process of mixing involves the irreversible transformation of initially segregated phases into a homogeneous state \citep[\textit{e.g.}][]{v2019}. Unlike confined mixtures, a state of complete homogeneity can never be achieved in an expanding turbulent thermal due to the continuous entrainment of the surrounding fluid. The competition between homogenization, driven by local stirring motions, and the production of compositional heterogeneities through the large-scale entrainment of surrounding fluid determines the degree of mixing within a thermal. It remains an open question whether these two processes are in dynamic equilibrium.

To our knowledge, the statistical distribution of the concentration or temperature field in a turbulent thermal has not yet been investigated. Numerical simulations of well-developed turbulent thermals \citep{lzlwa2015,lj2019,mjl2020,oc2020,vr2022,mjy2022,mm2023,mjlp2023,m2025} have predominantly investigated their global behavior in terms of the effects of moisture content on entrainment rates or the influence of stratification on thermal rise. Previous experimental work has also examined concentration measurements in puffs \citep{yckcbb1994,gj2007}, plumes \citep{yckcbb1995}, and thermals \citep{bbr1984,mb1992,zlla2013}. However, they mostly provide average concentration measurements \citep{mb1992,bj2005,bj2010,zlla2013}. Here, we quantify the time evolution of the distribution of the concentration or temperature fields, as measured by their probability density function (PDF), using both laboratory experiments and numerical simulations at different Schmidt numbers $Sc=\nu/D$. The main contribution of this work is to investigate the bulk concentration distribution and show that it becomes self-similar and is well described by a truncated exponential PDF, which provides a simple model for the finite-order concentration moments. Our direct 3D numerical simulations match the experimental Reynolds number but use a significantly lower Schmidt number ($Sc=1$). We analyze the influence of $Re$ and $Sc$ on the evolution and distribution of the concentration field. 

The paper is organized as follows. We first recall the dimensional analysis of turbulent thermals. We propose predictions for the time evolution of the moments of the concentration field. In Section~\ref{sec:thermal_exp_experiment}, we describe the experimental setup and numerical simulations. Section~\ref{sec:qualitative} provides a qualitative description of turbulent thermals in experiments and simulations, including the definition of thermal boundaries and their dependence on concentration and vorticity thresholds. Section~\ref{section:EvolutionPDF} presents a quantitative analysis of concentration distributions and moments, which suggests that the concentration field follows an exponential distribution far from the source. Finally, we discuss the effects of diffusivity ($Sc$) on concentration evolution in Section~\ref{subsec:Sc}.

\section{Dimensional analysis}
\label{sectionDimensionalAnalysis}

\subsection{Bulk properties}
\label{subsection:BulkProperties}

Consider a turbulent thermal formed by the release of a volume $V_0$ of a dense fluid, with density $\rho_a+\Delta\rho_0$, into an ambient fluid of density $\rho_a$. We denote by $d=(6 V_0/\pi)^{1/3}$ the equivalent diameter at release (\textit{i.e.} the diameter of a sphere of volume $V_0$). We assume that the density difference arises from fluctuations in concentration, but the following considerations also apply to thermally induced density variations, provided the Boussinesq approximation remains valid (\textit{i.e.} small relative density variations, and negligible dissipative heating and compressibility effects). We denote by $c_0$ the initial concentration at release, in kilograms of solute per unit volume, and by $c(\mathbf{x},t)$ the concentration at position $\mathbf{x}$ and time $t$. $D$ is the diffusivity of the concentration field, and $\nu$ is the kinematic viscosity, which we assume to be independent of concentration.

Previous experiments have shown that a turbulent thermal continuously entrains ambient fluid as it falls or rises \citep{b1954,s1957,t1979,zlla2013}. As a result, the volume $V$ of the thermal increases over time, while both its average concentration $\langle c \rangle$ and mean density difference $\Delta \rho$ relative to the surrounding fluid decrease \citep[\textit{e.g.}][]{b1954,mtt1956}.
Since the loss of material from the thermal is small \citep{lj2019}, we assume that the mass of solute within the thermal, defined as $\mathcal{M} = \langle c \rangle V$, is conserved and equal to its initial value
\begin{equation}
 \mathcal{M}_0 = c_0 V_0.
\end{equation}
This implies that the total buoyancy $\mathcal{B}=\frac{\Delta \rho}{\rho_a} g V$ of the thermal is constant and equal to its initial value $\mathcal{B}_0$ as defined by equation \ref{eq:InitialBuoyancy}.

In the Boussinesq approximation limit, a thermal is fully defined by $V_0$, $\mathcal{B}_0$, $\mathcal{M}_0$, and by the transport properties $\nu$ and $D$. As shown by \cite{b1954}, dimensional analysis provides valuable predictions for the bulk properties of a thermal. For instance, the vertical position $z_{c}$ of the center of mass of the thermal is a function of $t$, $V_0$, $\mathcal{B}_0$, $\mathcal{M}_0$, $\nu$, and $D$. This gives a set of seven quantities with three independent units. According to Vaschy-Buckingham's theorem, only four independent dimensionless quantities can be formed from these quantities, one possible set being
\begin{equation}
 \frac{z_c}{\mathcal{B}_{0}^{\frac{1}{4}} t^{\frac{1}{2}}}, \quad \frac{t}{t_g}, \quad Re = \frac{\mathcal{B}_{0}^{\frac{1}{2}}}{\nu}, \quad Pe = \frac{\mathcal{B}_{0}^{\frac{1}{2}}}{D}, \label{eq:AD1}
\end{equation}
where $t_{g}$ is a free-fall timescale defined as
\begin{equation}
t_{g} = \sqrt{\frac{\rho_a}{\Delta\rho_{0}}\frac{d}{g}} = \sqrt{\frac{\pi}{6}} \frac{d^{2}}{\mathcal{B}_0^{\frac{1}{2}}}.
\end{equation}

This implies that $z_{c} \mathcal{B}_{0}^{-1/4} t^{-\frac{1}{2}}$ is a function of $t/t_{g}$, $Re$ and $Pe$, or, equivalently, that
\begin{equation}
z_{c} = \mathcal{B}_{0}^{\frac{1}{4}} t^{\frac{1}{2}} f_z(t/t_g,Re,Pe) \label{eq:z_1} ,
\end{equation}
where $f_z$ is an unknown non-dimensional function. 
Similar analyses for the radius $r$ of the thermal, its vertical velocity $w$, and its average concentration $\langle c \rangle$ yield 
\begin{align}
 r &= \mathcal{B}_{0}^{\frac{1}{4}} t^{\frac{1}{2}} f_r(t/t_g,Re,Pe), \label{eq:r_1} \\
 w &= \mathcal{B}_{0}^{\frac{1}{4}} t^{-\frac{1}{2}} f_w(t/t_g,Re,Pe), \\
 \langle c\rangle &= \mathcal{M}_0 \mathcal{B}_{0}^{-\frac{3}{4}} t^{-\frac{3}{2}} f_c(t/t_g,Re,Pe),
\end{align}
where $f_r$, $f_w$, and $f_c$ are unknown non-dimensional functions. 
The prediction for $\langle c \rangle$ can also be derived from mass conservation $\langle c\rangle = c_0 V_0/V \sim c_{0} V_{0}/r^{3}$, using \eqref{eq:r_1} to express $r$.

In these equations, $V_0$, $\mathcal{B}_{0}$, $\mathcal{M}_0$, and $Re$, $Pe$ are conserved, time-independent quantities that do not depend on the release conditions of the dense fluid. In contrast, $t_g$ is influenced by the specifics of the initial conditions. 
When far enough from the release point, the thermal is expected to become independent of its initial conditions \citep{b1954}. In this limit, $r$, $w$, and $\langle c\rangle$ become independent of $t_{g}$.  In addition, laboratory experiments and field observations of atmospheric thermals \citep[\textit{e.g.}][]{w1958,w1959} and volcanic clouds \citep[\textit{e.g.}][]{t2007} have shown no measurable effect of $Re$ and $Pe$ on the time evolution of thermal size and velocity.
When $t \gg t_{g}$, these assumptions imply that \citep{b1954,mtt1956}
\begin{align}
 z_c &\sim \mathcal{B}_{0}^{\frac{1}{4}} t^{\frac{1}{2}} , \label{eq:z_2} \\
 r &\sim \mathcal{B}_{0}^{\frac{1}{4}} t^{\frac{1}{2}} , \label{eq:r_2} \\
 w &\sim \mathcal{B}_{0}^{\frac{1}{4}} t^{-\frac{1}{2}}, \label{eq:w_2} \\
 \langle c\rangle &\sim \mathcal{M}_0 \mathcal{B}_{0}^{-\frac{3}{4}} t^{-\frac{3}{2}}. \label{eq:c_2}
\end{align}
Combining equations \eqref{eq:z_2} and \eqref{eq:r_2} implies that the radius of the thermal is proportional to the distance it has traveled, which gives
\begin{equation}
r = \alpha z_c, 
\label{eq:radius depth}
\end{equation}
where $\alpha$ is the \textit{entrainment coefficient} \citep{mtt1956}.

\subsection{Moments of concentration}
\label{subsection:DimensionalAnalysisMoments}

Using dimensional analysis along the same lines as in Section \ref{subsection:BulkProperties}, the moments of order $k$ $\langle c^k \rangle$ of the concentration field can be expressed as
\begin{equation}
\langle c^k \rangle = \left(\mathcal{M}_{0} \mathcal{B}_{0}^{-\frac{3}{4}} t^{-\frac{3}{2} }\right)^k f_{c,k}(t/t_g,Re,Pe),
\label{eq:moments_c_1}
\end{equation}
where $f_{c,k}$ are unknown dimensionless functions. 

As explained in section \ref{subsection:BulkProperties}, we expect that the thermal becomes independent of its initial condition when far enough from the release point \citep{b1954}.
In this limit, when $t\gg t_g$, the moments become independent of $t_{g}$, which gives
\begin{equation}
\langle c^k \rangle = \left(\mathcal{M}_{0} \mathcal{B}_{0}^{-\frac{3}{4}} t^{-\frac{3}{2} }\right)^k f_{c,k}(Re,Pe).
\label{eq:moments_c_2}
\end{equation}
As a result, the moments of $c/\langle c \rangle$, expressed as 
\begin{equation}
 \langle (c/\langle c \rangle)^k \rangle = \frac{\langle c^k \rangle}{\langle c \rangle^{k}} = \frac{f_{c,k}(Re,Pe)}{f_{c,1}(Re,Pe)^{k}},
 \label{eq:moments_c_2.5}
\end{equation}
are predicted to be independent of time, which implies that the probability density function (PDF) of $c/\langle c \rangle$ is also independent of time.
Equivalently, this predicts that the PDF of $c$ is expected to become self-similar in time when $t\gg t_g$.
Since the moments of $c/\langle c \rangle$ depend on $t/t_g$ close to the source (Eq. \ref{eq:moments_c_1}), this prediction does not hold in that region.

We can also express the moments of concentration as
\begin{equation}
\frac{\langle c^k \rangle}{c_0^k} = \left(\frac{t}{t_g} \right)^{-\frac{3}{2} k} f_{c,k}(Re,Pe).
\label{eq:moments_c_3}
\end{equation}
This form has the disadvantage of artificially introducing parameters related to the source ($c_0$ and $t_g$) but will prove more convenient for comparison with experimental and numerical data (section \ref{section:EvolutionPDF}).

The dimensional analysis of the moments of concentration (Eq. \ref{eq:moments_c_2}) implies that the shape of the PDF of $c$ may vary with $Re$ and $Pe$. As a result, we aim to evaluate the predictions for the time evolution of the moments of $c$ and to investigate the effect of $Re$ and $Pe$ on the shape of its distribution.  
Although it has been convincingly argued that the bulk properties of the thermal are expected to be independent of viscosity and diffusivity \citep{t1946,b1954,s1957,mtt1956}, this has yet to be tested for the concentration distribution.

\section{Experiments and numerical simulations}
\label{sec:thermal_exp_experiment}
\subsection{Experiments}

\begin{figure}
 \centerline{\includegraphics[width=0.6\linewidth]{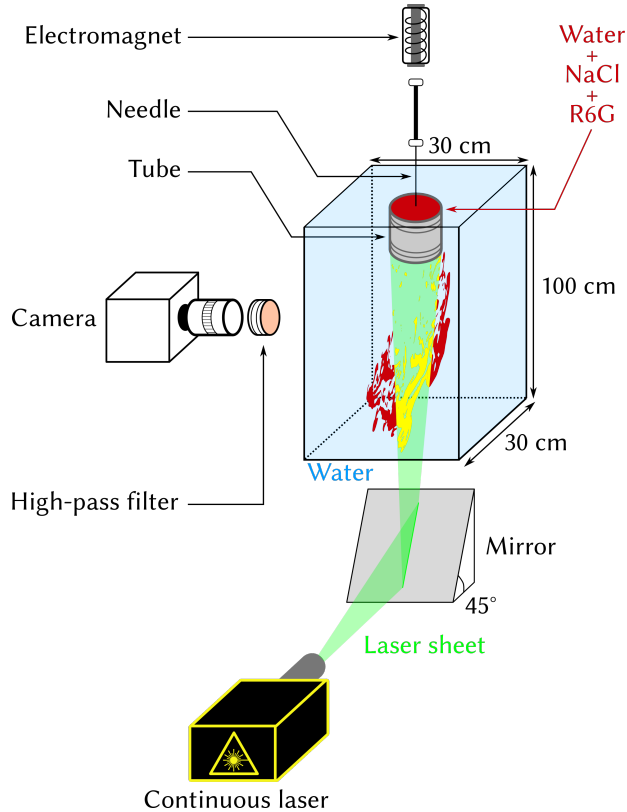}}
 \caption{Schematic of the experimental setup using laser-induced fluorescence on a turbulent thermal flow.}
 \label{fig:thermal_setup}
\end{figure}

In our experiments, we release a volume of an aqueous solution of sodium chloride (NaCl) into a $30 \times 30 \times 100$ cm tank containing fresh water (Fig.~\ref{fig:thermal_setup}). The NaCl solution is initially contained in cylindrical aluminum tubes with equal diameter and height $s_0 = 15$, $30$, $45$, and $60$ mm, corresponding to released volumes $V_0 = 2.7$, $21.2$, $71.6$, and $169.6$ mL, respectively. We use the diameter $d$ of the equivalent sphere of volume $V_0$, \textit{i.e.} $d=(3/2)^{1/3}s_0$, as a measure of the initial size of the released volume. We seal the tube at both ends with latex membranes stretched over the openings and proceed to fill it with the NaCl solution. We submerge the tube in the tank to a depth of approximately 15 cm below the surface, and we carefully pierce the upper membrane with a needle to release any trapped air bubbles. A negligible amount of fluid is drawn out of the tube during the operation due to the retraction of the membrane. After one minute at rest, we release a needle previously held in place by an electromagnet and abruptly puncture the lower membrane, which retracts rapidly. The fluid contained in the tube is quickly released due to the density difference with the ambient fluid. For the 15 mm tube, we pierce both membranes simultaneously, as, in this case, the retraction of the upper membrane draws a non-negligible amount of fluid out of the tube. As a result, dense fluid is released from the top of the tube, significantly altering the initial conditions of the fluid release.

We obtain the concentration field of a fluorescent dye (Rhodamine 6G) initially mixed with the NaCl solution using Laser-Induced Fluorescence (LIF). We illuminate the tank from below with a vertical laser sheet (532 nm) produced by a continuous 400 mW Nd:YAG laser coupled with a cylindrical lens (Fig.~\ref{fig:thermal_setup}). We image the LIF field using a PCO Edge 5.5 camera ($2560 \times 2160$ pixels, 20 Hz, 16-bit dynamic range) fitted with a Nikon Nikkor 50 mm f/1.2 lens. We use a high-pass filter ($> 540$ nm) to remove the laser wavelength and isolate the emission signal of Rhodamine 6G (between 500 and 700 nm). We spatially calibrate the images using a grid, remove the background, and correct for laser sheet heterogeneities. We obtain the concentration field using a calibration between light intensity and concentration (Appendix \ref{appA}). Finally, we have corrected the concentration of the dye absorption along each ray path \citep{crimaldi2008planar} (see more details in Appendix \ref{appA}). The experimental concentration values are limited by the dynamic range of the camera (16-bit resolution) and span at most five orders of magnitude in intensity. In the manuscript, the quantitative analysis is restricted to concentration values within the calibrated range and above the threshold used to define the thermal.

The field of view of the camera spans $32.8 \times 38.5$ cm (horizontal $\times$ vertical). When scaled by the equivalent diameter $d$, this corresponds to fields of view of $19 \times 22$, $10 \times 11$, $6 \times 7$, and $5 \times 6$ for tube diameters of 15, 30, 45, and 60 mm, respectively. The field of view starts at 5.3, 9.1, 8.3, and 7.6 cm below the bottom of the tube, which, when made dimensionless, gives 3.1, 2.6, 1.3, and 1.1 times the equivalent diameter.

Our experiments are divided into four categories, (A) to (D), depending on the tube used to release the dense fluid (Table \ref{tab:thermal_experiments}). For each tube, we choose the NaCl concentration such that the Reynolds number is maximized while keeping a concentration low enough to prevent significant optical distortion when the thermal enters the field of view of the camera. We measure the density of the NaCl solutions using an Anton Paar 35 Basic densitometer, which has an accuracy of approximately $0.1\%$. We estimate viscosities and refractive indices from data compiled by \citet{h2014crc}.

Table \ref{tab:thermal_experiments} gives the ratio $\Delta\rho_0/\rho_a$ selected for the four configurations corresponding to different tube diameters $s_0$ and equivalent sphere diameter $d$. The range of initial density differences corresponds to a range of total buoyancy $\mathcal{B}_0$ and translates into a viscosity range of $\mu_0=1.03 - 1.60~\si{\milli\pascal\second}$ with kinematic viscosity between $1.02\times 10^{-6}$ and $1.38\times 10^{-6}$~\si{\meter\squared\per\second}) and a refractive index range of $1.336-1.370$. In comparison, the viscosity of fresh water is $\mu_a=1.005~\si{\milli\pascal\second}$ and its refractive index is $1.333$. However, it should be noted that the cloud formed by the NaCl solution is significantly diluted when it enters the field of view of the camera, resulting in variations of viscosity and refractive index that are significantly smaller than those suggested by the values mentioned above. The maximum viscosity corresponding to the maximum salt concentration of the thermal (see Fig.~\ref{fig:meanmaxconcentration}b) is up to 3\% larger than the viscosity of water (1.6\%, 2.3\%, 3\%, 3\% for A, B, C, D cases). The average viscosity in the diluted thermal when it enters the field of view is close to that of water. Therefore, we neglect the effect of viscosity on the long-term evolution of the thermal and calculate the $Re$ using the viscosity of the ambient fluid (see Eq.~\ref{eq:AD1}).

We calculate $Pe$ from equation \eqref{eq:AD1} using the diffusivity of Rhodamine 6G in water, estimated as $D^{rhod}=\num{4.0e-10}~\si{\meter\squared\per\second}$ \citep{gaw2008}. The diffusivity of sodium chloride, $D^{salt}=\num{1.5e-9}~\si{\meter\squared\per\second}$, is approximately four times larger than the diffusivity of Rhodamine 6G. As a result, we expect the salt concentration field to differ from the Rhodamine 6G concentration field measured with LIF at the smallest scales. However, the two diffusivities yield large Schmidt numbers, $Sc=2500$ and $Sc=670$, respectively.

In each configuration, we conducted and analyzed three experiments to estimate flow variability. The unsteady nature of the flow, as well as the coupling between buoyancy and mean flow, makes the early evolution of the thermal sensitive to the initial conditions resulting from puncturing the membrane that seals the tube. The lower membrane does not always retract in the same way, thereby affecting the initial conditions. The measured concentration is normalized by the initial concentration $c_0=c(t=0)$ in the tube. In Section \ref{section:EvolutionPDF}, we show the evolution of the concentration moments and PDFs averaged over the three experiments for each Reynolds number.
\begin{table}
\begin{center}
 \begin{tabular}{cccccccc}
 \# & $s_0$ (mm)  & $d$ (mm) & $\Delta\rho_0/\rho_a$ & $\mathcal{B}_{0}$ ($\si{\meter^4\per\second\squared}$) & $\mu_0/\mu_a$ & $Re$ & $Pe$ \\
 (A) & $15 $ & $17.1 $ & $0.156$ & $\num{4.1e-6}$ & $1.62$ & $2012$ & $\num{5.04e6}$\\
 (B) & $45 $ & $51.5$ & $0.017$ & $\num{1.2e-5}$ & $1.05$ & $3452$ & $\num{8.64e6}$\\
 (C) & $30 $ & $34.3$ & $0.084$ & $\num{1.8e-5}$ & $1.25$ & $4175$ & $\num{1.04e7}$\\
 (D) & $60 $ & $68.7$ & $0.013$ & $\num{2.2e-5}$ & $1.04$ & $4646$ & $\num{1.16e7}$\\
 \end{tabular}
\end{center}
 \caption{Parameters used in the four types of experiments (A, B, C, and D) with $Sc = 2500$}
 \label{tab:thermal_experiments}
\end{table}

\subsection{Numerical simulations}
We conducted three-dimensional simulations of thermals with the open-source pseudospectral code Dedalus \citep{bvolb2016,bvolb2020}, using the setup described by \cite{lj2019}. The equations of conservation of momentum and concentration are solved under the Boussinesq approximation. Details about the simulations can be found in \citep{lj2019}; we follow the same approach. While experiments have a camera-limited range of concentration values (5 orders of magnitude), the numerical simulations resolve concentration values over eight orders of magnitude. The thermal is initialized at $t=0$ by introducing a spherical concentration perturbation of diameter $d$ into an otherwise homogeneous fluid. The spatial domain extends horizontally over $10d$ in both the $x$ and $y$ directions, and vertically over $20d$ in the $z$ direction. It is discretized in $256 \times 256 \times 512$ modes.
We benchmark our simulations against a simulation with a higher resolution of $512 \times 512 \times 1024$, kindly provided by Daniel Lecoanet (see simulation at $Re = 6300$ in \citep{lj2019}). At a given Reynolds number ($Re \approx 4631$), we find that the kinetic energy spectrum and PDF of concentration are similar between the two resolutions (see Fig.~12 in Supplementary Materials \citep{hldhl2025}).

The numerical simulations use Reynolds numbers that match those of the experiments. In contrast, the Schmidt number $Sc$ is set to 1 in all numerical simulations (for computational resource reasons), resulting in a concentration field that diffuses much faster than in the experiments, where $Sc=2500$. Consequently, the Péclet number is also a factor $2500$ larger in the experiments than in the simulations. This allows us to investigate the effect of the Schmidt number on concentration distributions.

Note that turbulent thermals are at first order cylindrically symmetric. Therefore, we analyze two orthogonal two-dimensional planes from the 3D simulations, as it would be similar to the laser plane of the experiments. In Section \ref{section:EvolutionPDF}, we will show the evolution of concentration moments and PDF for the average of these two orthogonal planes and compare them to the experiments performed at the same Reynolds numbers.

\section{Qualitative description of a turbulent thermal}
\label{sec:qualitative}
\subsection{Macroscopic evolution}
\label{subsec:macro}

\begin{figure} 
\centerline{\includegraphics[scale=1]{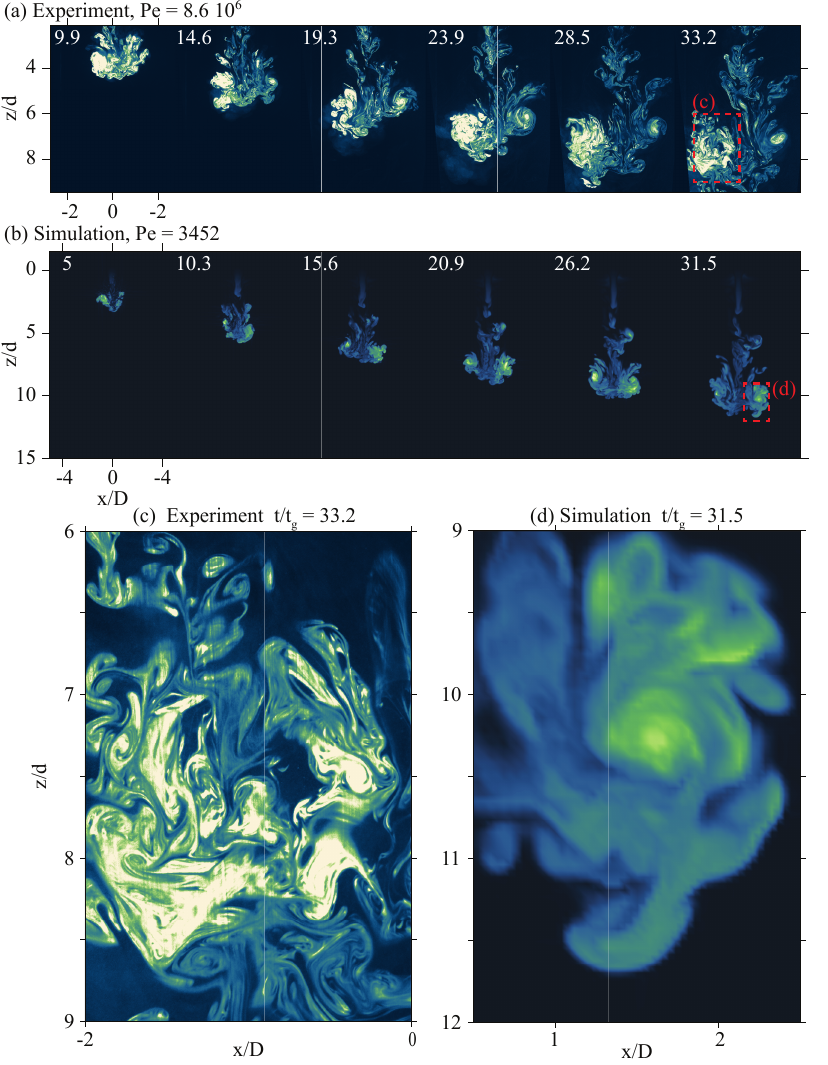}}
\caption{(a,b) Evolution of a turbulent thermal for one experiment with $Re = 3452$ ($Pe=8.64 \times 10^6$) and one simulation with $Re=3452$ ($Pe=3452$), respectively. Note that the concentration scale changes at each time step for better visualization. On each image, the time is normalized by $t_g$. (c,d) Close-up snapshots of the last time step for the experiment (a) and simulation (b).} 
 \label{fig:evolution_exp_sim}
\end{figure}

Fig.~\ref{fig:evolution_exp_sim} shows the evolution of the concentration field for a given $Re=3452$ for one experiment and a simulation. During its fall, the radius of the turbulent thermal increases as the surrounding liquid is entrained and incorporated into the thermal (Fig.~\ref{fig:evolution_exp_sim}a,b), a phenomenon referred to as turbulent entrainment. We observe the interaction between a large-scale engulfment, producing an entraining flow behind the thermal, and the development of small-scale turbulent fluctuations. As a result, turbulent entrainment progressively dilutes the fluid within the thermal, through stirring and mixing processes. In both experiments and simulations, we find that the radius of the thermal increases with depth, in agreement with the prediction of equation \ref{eq:radius depth} (see Fig.~1 in Supplementary Materials \citep{hldhl2025}). Our simulations have an entrainment coefficient $\alpha$ close to $0.18$, as observed in previous numerical and experimental studies \citep{lj2019,bj2010,lzlwa2015}. In the experiments, the entrainment  coefficient is larger, especially at the smallest Reynolds number experiments ($\alpha \sim 0.4$ for the smallest $Re$, $\alpha \sim 0.25$ for the others $Re$), but this may be due to the difference in the release mechanism. The entrainment coefficient is very sensitive to the initial conditions, and can only be estimated by averaging a large number of experiments \cite{zlla2013}.

Overall, both experiments and simulations evolve similarly at a large scale. The differences between the experiments and simulations in the depth of the thermals (at a given time) are due to the initial condition (release from a tube or from a sphere inside the surrounding fluid). While the Boussinesq approximation is used in the simulations, the experimental thermals experienced a large variation in density during the initial phase. In the close-up snapshots (Fig.~\ref{fig:evolution_exp_sim}c,d), we observe significant differences in the scale of the structures inside the turbulent thermal, which are probably due to the difference in Schmidt number (see discussion in Sec. \ref{subsec:Sc}).

\subsection{Anatomy of a turbulent thermal}
\label{subsec:anatomy}
In this section, we describe the anatomy of a turbulent thermal and investigate the concentration distribution in its different parts. We consider concentration fields from an experiment and a simulation at the same Reynolds number ($Re=4175$), and we focus on snapshots far from the source, \textit{i.e.} $t/t_g=51.25$ in the experiment and $t/t_g=31.6$ in the simulation. In the following, the concentration is normalized by the initial concentration $c_0$.

A buoyant thermal is composed of a head, containing most of the initial buoyant fluid, followed by a turbulent wake. The head has an ellipsoidal shape that is flattened in the vertical direction. It consists of a vortex ring, the so-called vortex cores of the thermal, and a peripheral region where the surrounding fluid is entrained in the thermal. The regions with high values of concentration are the undiluted cores \citep{bcj1988,rk2010}.

Although turbulent thermal structures are qualitatively well-defined, tracking their position, shape, and contour may be quantitatively challenging. In a miscible thermal, the entrainment of the surrounding fluid and the diffusion of buoyant fluid complicate the definition of its contour. Previous works implemented various methods, from an ellipsoidal shape defined by the horizontal and vertical average of the buoyancy field \citep{zlla2013} or using the velocity field \citep{rc2015,lj2019,mjl2020}. Overall, the envelope of the thermal depends on a threshold value of the buoyancy field (temperature, concentration, or density).

\begin{figure}
 \centerline{\includegraphics[scale=1]{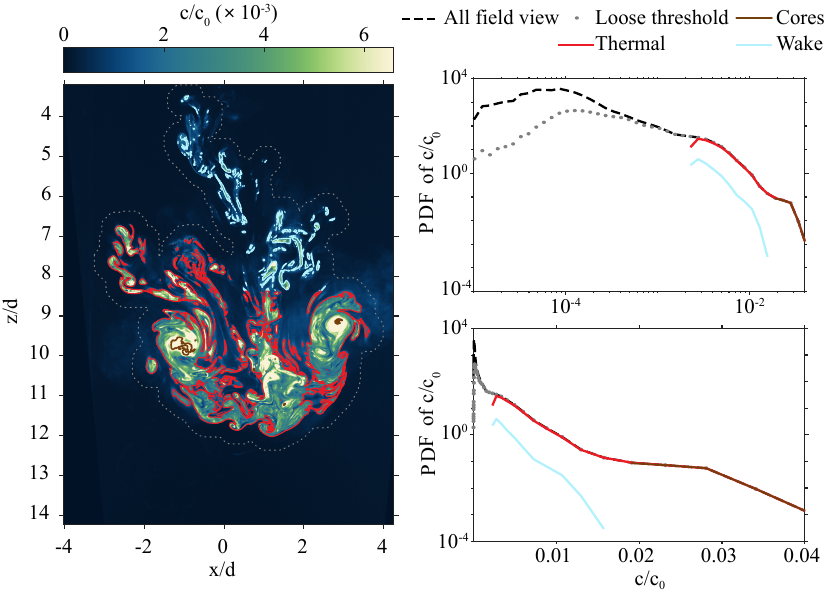}}
 \caption{(a) Snapshot of the concentration field at $t/t_g = 51.25$ for one of the three experiments at $Re = 4175$. Colored lines denote the mask applied to the snapshot to produce the PDF of the concentration in (b, c). (b) and (c) PDF of the concentration as a function of the concentration in log-log or semi-log, respectively. The dashed black line represents the PDF of the whole snapshot. The gray dotted line shows the PDF of the thermal using a loose threshold. The inside of the thermal comprises all pixels inside the cyan and red masks. Brown and cyan lines in (b,c) (and mask in (a)) denote the different parts of the thermal: wake and cores, respectively.}
 \label{fig:anatomy_exp}
\end{figure}

\begin{figure}
 \centerline{\includegraphics[scale=1]{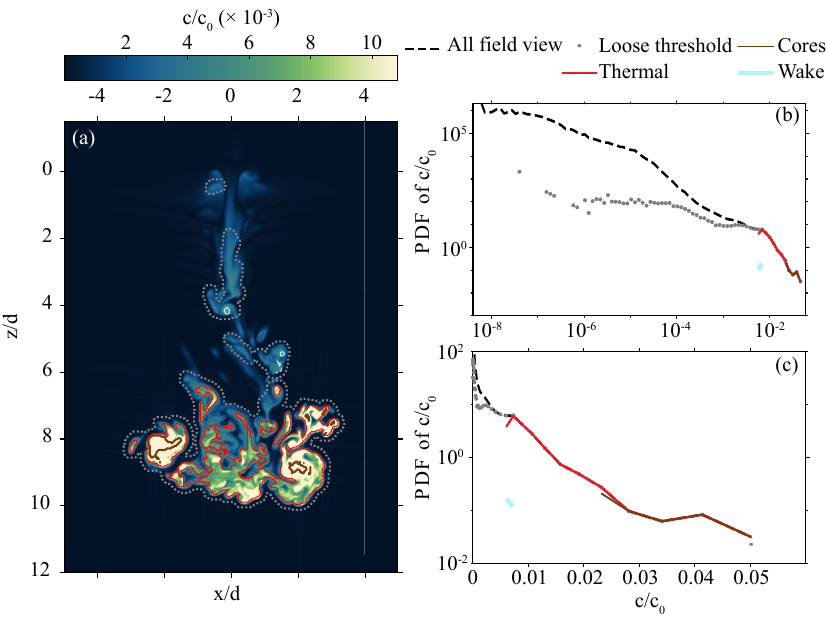}}
 \caption{(a) Snapshot of the concentration field at $t/t_g = 31.6$ for one of the simulations at $Re = 4175$. Colored lines denote the mask applied to the snapshot to produce the PDF of the concentration in (b, c). (b) and (c) PDF of the concentration as a function of the concentration in log-log or semi-log, respectively. The dashed black line represents the PDF of the whole snapshot. The gray dotted line shows the PDF of the thermal using a loose threshold. The inside of the thermal comprises all pixels inside the cyan and red masks. Brown and cyan lines in (b,c) (and mask in (a)) denote the different parts of the thermal: wake and cores, respectively.}
 \label{fig:anatomy_sim}
\end{figure}

Figs.~\ref{fig:anatomy_exp} and \ref{fig:anatomy_sim} show the effect of the definition of the boundary of the thermal on its spatial extent and the concentration distribution. We investigate two different ways of defining the limits of the thermal: a tight threshold corresponding to a concentration equal to a fixed fraction $0.5$ of $\langle c\rangle$, the average of $c$ within the thermal defined by the mask, and a loose definition of the thermal limit corresponding to the same threshold followed by a dilation operation using the nearest neighbors, which expands the mask by adding surrounding pixels. We have tested the effect of varying the concentration threshold on the normalized concentration distribution and find no significant change in the shape of the distribution at intermediate values of $c/\langle c\rangle$. In addition, the threshold used here is close to the experimental background level. The threshold value is the same for all experiments and simulations; it is constant in terms of the value of $c/\langle c \rangle$, but time-dependent in terms of $c$. The reasons for this choice are explained in Section \ref{subsec:PDFconcentration}. Both masks include concentration values of the head, cores, and wakes of the thermal. The wake is qualitatively determined by the material left behind the buoyant thermal by detrainment. For the tight threshold, we split the mask into two parts using a given value of $z/D$: the head of the thermal and the wake, which allows us to identify their respective PDFs. We also qualitatively define the cores by using the extreme values with the criterion $c/c_0> (60\%\,\text{of}\, max(c/c_0))$, which corresponds here to a change in slope in the PDF of $c/c_0$ (Fig.~\ref{fig:anatomy_exp}c). 

Fig.~\ref{fig:anatomy_exp}a shows a turbulent thermal with $Re=4175$ and the two masks. In Figs.~\ref{fig:anatomy_exp}b,c, we compare the probability density functions of the concentration values for the different regions in the thermal. Most of the concentration values are below $10^{-3}$ and correspond to the background values outside the thermal (Fig.~\ref{fig:anatomy_exp}b). The PDF of the whole field of view and both masks are very similar for concentration values above $10^{-3}$. The PDFs of the two parts of the tight mask, the head and the wake, have the same shape (Fig.~\ref{fig:anatomy_exp}b), except there are no extreme values of concentration in the wake. These extreme values mostly come from the cores inside the head, as already observed in previous experiments \citep{j1992}. Fig.~\ref{fig:anatomy_exp} shows that the loose mask includes values that are part of the background or surrounding fluid. Therefore, the shape of the PDF below $10^{-3}$ depends significantly on the image analysis, especially the number of pixels considered within the thermal.

In Fig.~\ref{fig:anatomy_sim}, we have applied the same threshold and mask to a simulation (with the same Reynolds number as in Fig.~\ref{fig:anatomy_exp}). For a concentration value above $10^{-2}$, the PDF values of loose and tight masks are very similar. Most of the high values of concentration correspond to the cores of the thermal inside the head (Fig.~\ref{fig:anatomy_sim}b). In Fig.~\ref{fig:anatomy_sim}b, the PDF of the smallest values (between $10^{-8}$ and $10^{-5}$) decreases as $1/c$, which is a characteristic of the initial perturbation at $t/t_g=0$. The values of concentration between $10^{-5}$ and $10^{-2}$ surround the thermal defined by the tight mask. The shape of the PDF for these values depends significantly on the image processing used to define the mask. 

Based on Fig.~\ref{fig:anatomy_exp} and \ref{fig:anatomy_sim}, we have chosen to analyze our data using the tight threshold. Fig.~\ref{fig:anatomy_exp}c and \ref{fig:anatomy_sim}c show semi-log versions of the PDF of concentration, which highlights its shape for the values above $10^{-3}$. It suggests that the PDF of $c$ decreases exponentially at intermediate values of $c$, between the threshold and an upper value, which is $\simeq 0.013$ in the experiment (Fig.~\ref{fig:anatomy_exp}b) and  $\simeq 0.03$ in the simulation (Fig.~\ref{fig:anatomy_sim}b) snapshots. Figs.~\ref{fig:anatomy_exp}c and \ref{fig:anatomy_sim}c suggest that the deviations from the exponential PDF at high concentrations can be attributed to the cores.

\section{Evolution of PDF and moments of concentration}
\label{section:EvolutionPDF}

In this section, we investigate the moments and PDFs of the concentration. We will examine the time dependence from the release of the buoyant fluid (for the simulations) to large distances and long times after the release ($t/t_g>10$). We will also compare the differences and similarities between experiments and simulations that have been performed at the same Reynolds numbers but different Péclet numbers.

\subsection{Moments of concentration}
\label{subsec:moments}
\begin{figure}
 \centerline{\includegraphics[scale=1]{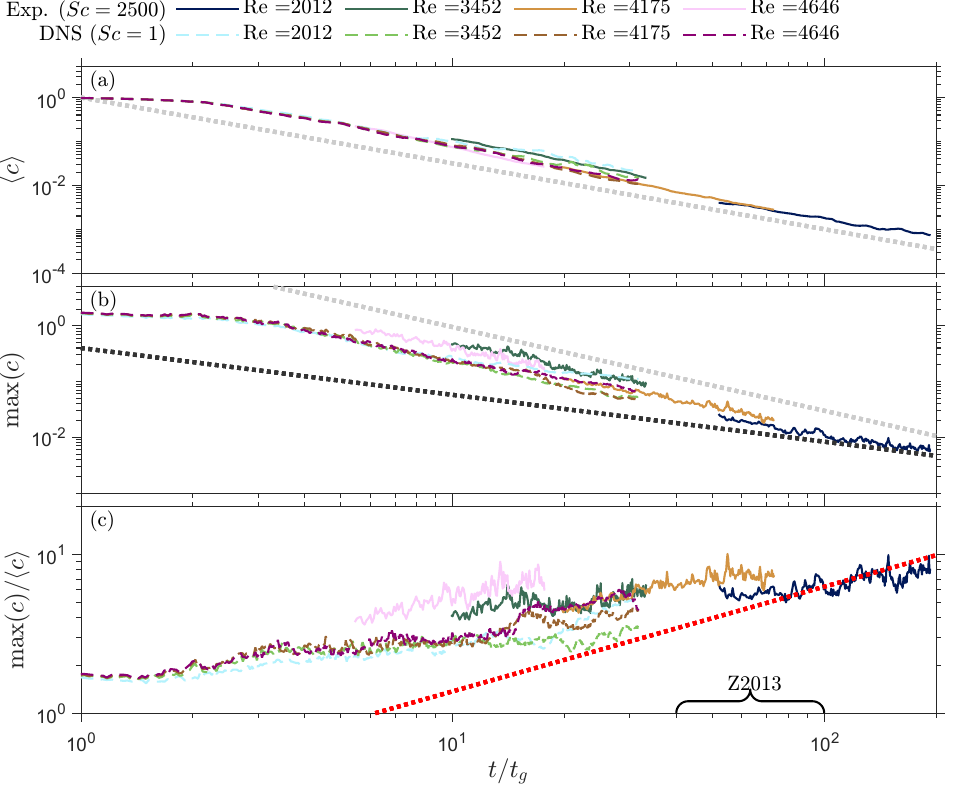}}
 \caption{(a) Mean of concentration $c/c_0$ as a function of time $t/t_g$ for the average of three experiments at each $Re$ (solid lines) and the average of two simulations at each $Re$ (dashed lines). (b) Maximum concentration $c/c_0$ as a function of time $t/t_g$. In (a) and (b), the gray and black dotted lines show the power laws $t^{-3/2}$ and $t^{-0.84}$ as in \cite{zlla2013}. (c) The ratio between the maximum concentration and its mean as a function of time $t/t_g$. The red dashed line shows the ratio of the two previous power laws. Results of \cite{zlla2013} (Z2013) span only between 40 and 100 $t/t_g$. Note that all power laws intend to show the slope, i.e., they have an arbitrary pre-factor.}
 \label{fig:meanmaxconcentration}
\end{figure}

Fig.~\ref{fig:meanmaxconcentration}a,b shows the mean concentration $\langle c/c_0 \rangle$ and the maximum concentration as functions of time for all experiments and simulations. These are calculated from $t=0$ in the simulations, but we only focus on $t/t_g > 1$. At larger $t/t_g$, the mean decreases following a power law close to $t^{-3/2}$, in agreement with the prediction of dimensional analysis (Eq.~(\ref{eq:moments_c_3})). We have checked the sensitivity of the evolution of the maximum concentration to the resolution. Since the concentration is averaged within each pixel, the absolute value of the maximum depends on the resolution of the concentration structures inside the thermal. In addition, the number of pixels covering the thermal is not constant, as the thermal radius increases as the square root of time (about 5 times more pixels per thermal radius at $t/t_g=30$ than at $t/t_g=0$). By quantifying this effect by degrading the saved concentration fields a posteriori, \textit{i.e.,} while maintaining a constant number of pixels per thermal radius, we conclude that the decrease of the maximum concentration is significant and not a geometrical artifact. For both experiments and simulations, the maximum concentration decreases more slowly than the mean concentration. The exponents of the maximum and mean concentrations are similar between our experiments and simulations. However, the experiments can be offset in time from one another. This is due to the inherent variability between each experiment in the initial condition, which determines the value of $t_g$ and the time origin. We also found that these slopes agree reasonably well with the power-law fit from previous experiments for a similar Reynolds number \citep{zlla2013}. Note that the fit from \citep{zlla2013} is obtained in a short time range, which can introduce non-negligible errors in the exponent. Fig.~\ref{fig:meanmaxconcentration}c shows that the ratio between the maximum and the instantaneous concentration mean increases with time, which is consistent with the thermal cores evolving separately from the bulk of the thermal. The maximum concentration relative to the mean concentration in the simulations is smaller than in the experiments, which can be interpreted as the effect of a higher diffusion rate in the simulation.

\begin{figure}
 \centerline{\includegraphics[scale=1]{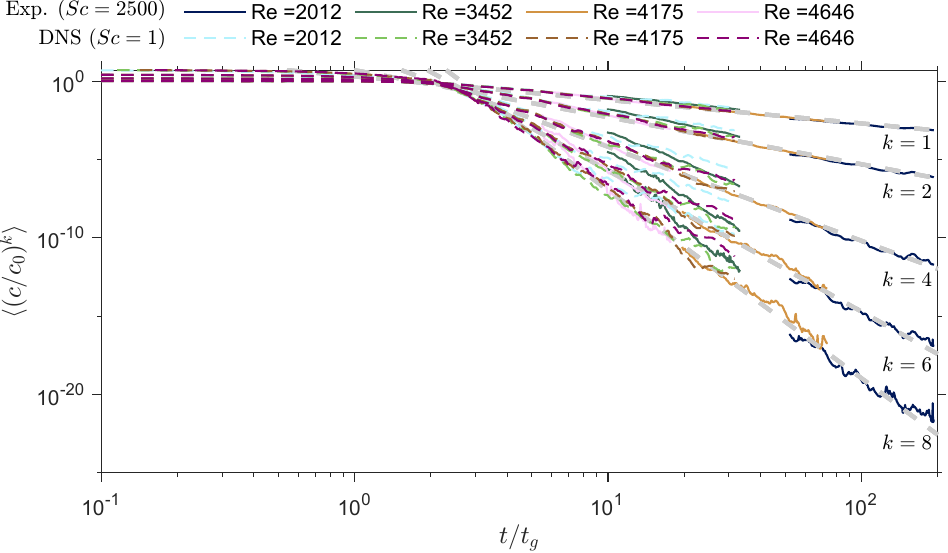}}
 \caption{Moments of concentration $c^k,\,\text{with}\, k=1,2,4,6,8$ as a function of time $t/t_g$ for the average of three experiments at each $Re$ (solid lines) and the average of two simulations at each $Re$ (dashed lines). The gray dotted lines show the moments as predicted by dimensional analysis and the exponential shape of the PDF (Eq.~\ref{eq:moment_concentration}).}
 \label{fig:momentconcentration}
\end{figure}

Fig.~\ref{fig:momentconcentration} shows the $k$-th moments of the concentration ($k=1$, 2, 4, 6, 8) over time. For each Reynolds number, we average the three experiments and two orthogonal planes in each simulation. 
The dashed gray lines show the predictions of the dimensional analysis (Eq.~\ref{eq:moments_c_3}) for the time evolution of the moments, $\langle c^k \rangle \sim \left(\frac{c_1 t}{t_g} \right)^{-\frac{3}{2} k}$. The values of the prefactors ($c_1=20$) and the implications for the shape of the PDF will be discussed in subsection \ref{subsec:PDFconcentration}. The time evolution of the moments is in good agreement with the predictions from dimensional analysis. Although the 8th moment of concentration decreases by 20 orders of magnitude over the investigated time interval, Eq.~\ref{eq:moments_c_3} still accurately predicts its evolution.
\subsection{PDFs of concentration}
\label{subsec:PDFconcentration}
\begin{figure}
\centerline{\includegraphics[scale=1]{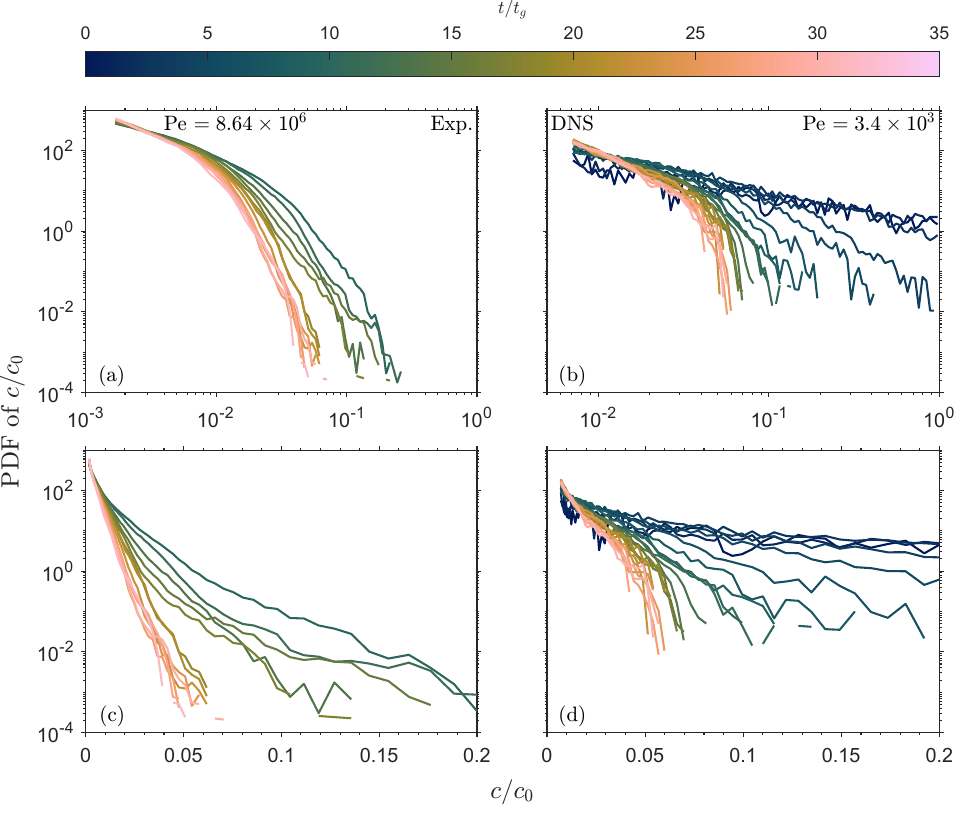}}
 \caption{PDF of $c$ as a function of $c$ and time $t/t_g$ for an experiment with $Re=3452$ (a,c) ($Pe=8.64 \times 10^6$) and a simulation with the same $Re$ (b,d) ($Pe=3.4 \times 10^3$). The top and bottom rows are log-log and semi-log plots, respectively.}
 \label{fig:PDFcexpsim}
\end{figure}

We have calculated the probability density function of the concentration as a function of time for all the experiments and simulations. The bins of the PDF are logarithmically spaced with a width in log-space of $0.066$ for the PDF of concentration $c/c_0$ and with a width in log-space of $0.12$ for the PDF of the concentration normalized by the mean concentration $\langle c \rangle$. Fig.~\ref{fig:PDFcexpsim} represents these PDFs for a given experiment and a simulation, with the same $Re=3452$. The PDFs of $c$ for all other Reynolds numbers are shown in the Supplementary Materials \citep{hldhl2025}. 
In Fig.~\ref{fig:PDFcexpsim}, the first PDF that is plotted corresponds to $t/t_g\sim 10$ for the experiment, and $t/t_g=0$ for the simulation, for which the concentration field is available from the start. In both cases, the PDFs of concentration exhibit a drift toward smaller concentration values, in agreement with the time evolution of $\langle c \rangle$ and $\mathrm{max}(c)$ (see figure \ref{fig:meanmaxconcentration}a,b). 
For $t/t_g > 10$, the shape of the PDF is relatively constant with time and qualitatively consistent between the simulations and the experiments. 

Fig.~\ref{fig:momentconcentration} confirms the prediction of the dimensional analysis that the moments evolve as $\sim t^{-\frac{3}{2}k}$.
Following the discussion of section \ref{subsection:DimensionalAnalysisMoments}, one should therefore expect the PDF of $c/\langle c \rangle$ to be independent of time. Fig.~\ref{fig:PDFc_meanc_expsim} shows the PDFs of $c/\langle c \rangle$ on log-log and semi-log axes for an experiment and a simulation, with $Re=3452$. Once normalized by $\langle c \rangle$, the PDFs of concentration in both the simulation and the experiment collapse (Fig.~\ref{fig:PDFc_meanc_expsim}a,b), in agreement with the prediction of section \ref{subsection:DimensionalAnalysisMoments}. 

A similar behavior is observed for all Reynolds numbers (see Figs.~5-7 in Supplementary Material \citep{hldhl2025}). In the simulations, the PDF initially exhibits a different shape for the first few $t/t_g$ because the thermal is not yet fully developed. In addition, the concentration PDF remains nearly identical between simulations with the same Reynolds number but different resolutions (see Fig.~12 in Supplementary Material \citep{hldhl2025}).

\subsection{Shape of the PDF of concentration}
\label{subsec:shapePDFconcentration}

\begin{figure}
\centerline {\includegraphics[scale=1]{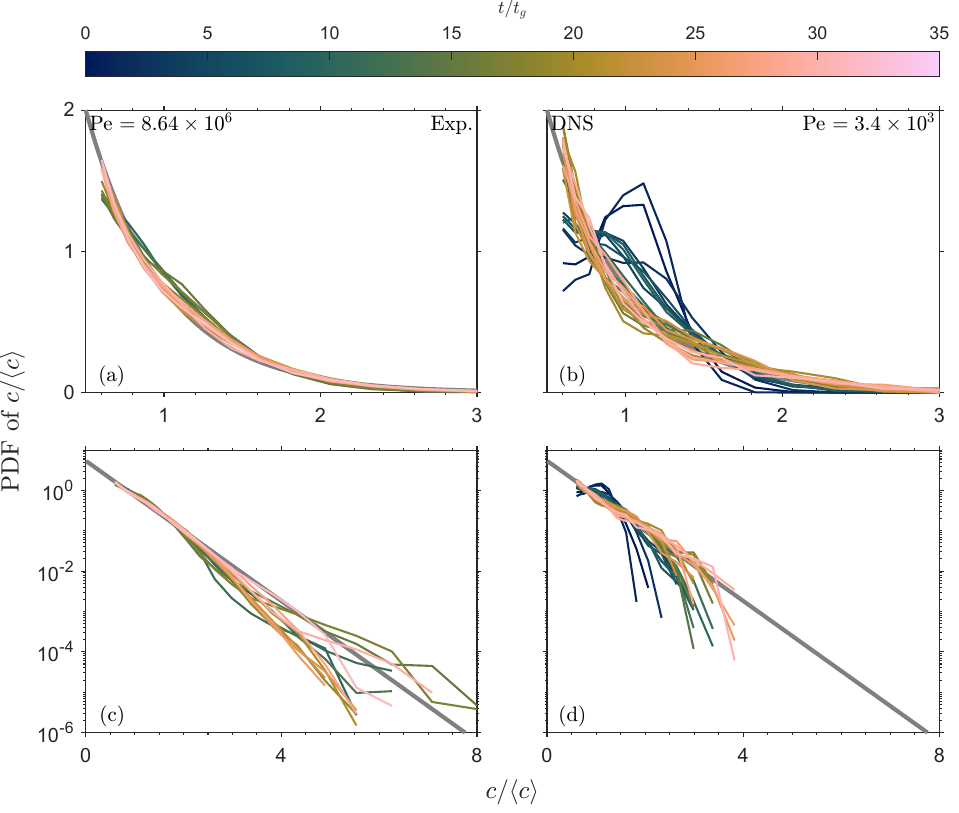}}
 \caption{PDF of $c/\langle c \rangle$ as a function of $c/\langle c \rangle$ and time $t/t_g$ for an experiment with $Re=3452$ (a, c) ($Pe=8.64 \times 10^6$) and a simulation at the same $Re$ (b, d) ($Pe=3.4 \times 10^3$). The top and bottom rows are linear-linear and semi-log plots, respectively. The gray lines show the exponential distribution  defined by Eq. \ref{eq:exponential_distribution}.}
 \label{fig:PDFc_meanc_expsim}
\end{figure}

Fig.~\ref{fig:PDFc_meanc_expsim} shows that the PDF has an exponential shape for intermediate values of $c/\langle c \rangle$ and for $(t/t_g>15)$ in both the experiment and the simulation. In the numerical simulations, the PDFs deviate significantly from the exponential at early times $(t/t_g<10)$, corresponding to the acceleration phase where self-similarity is not yet established. Similar deviations are expected in the experiments; however, they are not observed here as the experimental field of view was focused on the distal behavior far from the source, excluding the initial acceleration. Note that care must be taken when characterizing the underlying PDF. Assuming it is of the form $p(X=c /\langle c\rangle)=\beta \exp{(-\lambda X)}$ as suggested by our data, the parameters $\lambda$ and $\beta$ are fixed by the requirement that the integral of $p(c/\langle c \rangle)$ over the concentration domain is 1, and that the expected value (i.e., first moment) of $c/\langle c \rangle$ is 1.
If $c \in [0,\infty [$, then $\lambda=\beta=1$ and, if exponential, the PDF of $c /\langle c\rangle$ is
\begin{equation}
 p(X=c /\langle c\rangle)= \exp(-X).
\end{equation}
However, the PDFs we have calculated are bounded below by the threshold value used to define the thermal, which is given by a fixed value $\xi$ of the ratio between the concentration and the mean. Imposing that $\int_\xi^\infty p(X=c /\langle c\rangle) dX=1$ and $\int_\xi^\infty p(X=c /\langle c\rangle  )X dX=1$, we find that, if exponential, the truncated PDF of $c /\langle c\rangle$ is
\begin{equation}
 p(X=c /\langle c\rangle)|_{X>\xi}=\frac{1}{1-\xi} \exp\left(-\frac{X-\xi}{1-\xi} \right).
 \label{eq:exponential_distribution}
\end{equation}
The slope (in a semilogarithmic graph) of the apparent PDF is therefore dependent on the chosen threshold, and the effect is significant if the threshold $\xi$ is not $\ll 1$, which is the case here ($\xi=0.5$, see Section~\ref{subsec:anatomy}).
One implication is that, for meaningful comparisons, PDFs from different experiments, simulations, and times must be calculated over the same domain of $c/\langle c \rangle(t)$ (\textit{i.e.} same value of $\xi$). 

As discussed in section~\ref{subsec:anatomy}, this threshold is set to $\xi=0.5$ at all times for all experiments and simulations. The cumulative distribution function associated with \eqref{eq:exponential_distribution} is
\begin{equation}
 P(c/\langle c \rangle<X)|_{c/\langle c \rangle >\xi} = 1 - \exp\left(- \frac{X-\xi}{1-\xi} \right),
 \label{eq:exponential_cdf}
\end{equation}
and the moments are
\begin{equation}
\langle X^k \rangle = (1-\xi)^k \exp\left( \frac{\xi}{1-\xi} \right) \Gamma\left(k+1,\frac{\xi}{1-\xi}\right),
\label{eq:moment_truncated_exp}
\end{equation}
where the incomplete gamma law is given by
\begin{equation}
 \Gamma(k+1, \xi) = \int_{\xi}^{\infty} x^{k} e^{-x} dx.
 \label{eq:gammainc}
\end{equation}

Comparison of the experimentally and numerically determined PDF with the truncated exponential PDF given by \eqref{eq:exponential_distribution} (gray line in Fig.~\ref{fig:PDFc_meanc_expsim})
 shows that the distribution of $c/\langle c \rangle$ is well described by the exponential distribution \eqref{eq:exponential_distribution} for $c/\langle c \rangle$ smaller than 3 to 5, depending on $t$. According to equation \eqref{eq:exponential_cdf}, the probability of finding $c/\langle c \rangle < 3$ in the thermal is equal to $99.3$\%, which shows that the concentration in most of the thermal follows closely an exponential distribution, with the deviations at high concentrations being representative of less than 1\% of the thermal cross-section surface area. Using the empirical cumulative distribution function calculated from the concentration maps rather than equation \eqref{eq:exponential_cdf} gives similar results for experiments and simulations (see Figs.~8-11 in Supplementary Materials \citep{hldhl2025}).

\begin{figure}
\centerline {\includegraphics[scale=1]{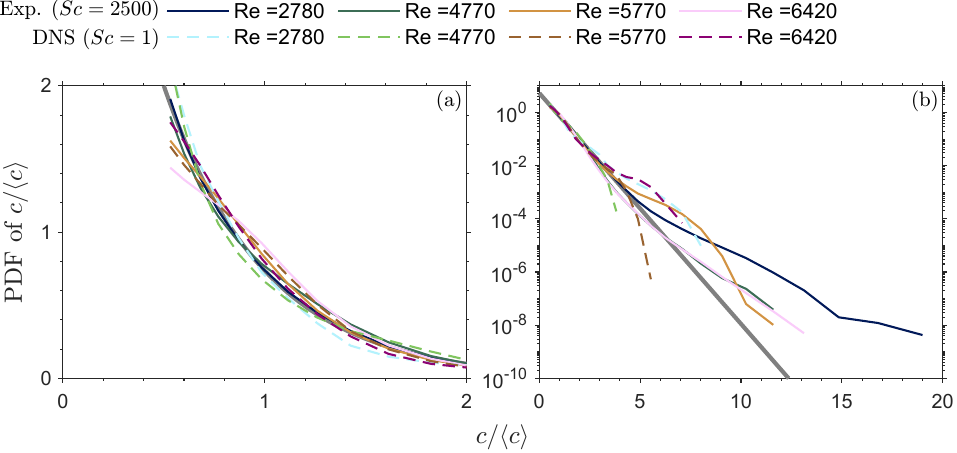}}
 \caption{PDF of time-averaged $c/\langle c \rangle$ as a function of $c/\langle c \rangle$ for all different $Re$ numbers for the experiments (solid lines) and the simulations (dashed lines). (a) and (b) are linear-linear and semi-log plots, respectively. The gray line is the truncated exponential distribution (Eq.~(\ref{eq:exponential_distribution})).}
 \label{fig:PDF_meanc}
\end{figure}

Once the PDFs of concentration have converged toward a self-similar distribution with time, we compare the time-averaged PDF and examine the effects of Reynolds or Péclet numbers (Fig.~\ref{fig:PDF_meanc}). 
The PDFs of $c /\langle c\rangle$ are averaged in time (for $t/t_g>10$) and averaged between runs for experiments (and between the two orthogonal planes for simulations). For intermediate concentration values ($c /\langle c\rangle<2$), the PDF slope seems to be independent of $Re$ or $Pe$ (Fig.~\ref{fig:PDF_meanc}a). For high concentration values ($c /\langle c\rangle>2$), there are significant deviations from the exponential decrease, but without any clear trend with Reynolds or Péclet number (Fig.~\ref{fig:PDF_meanc}b). These high values of concentration are related to the undiluted cores (see Section~\ref{subsec:anatomy}).\\

We can further evaluate our dimensional analysis and the exponential shape of the PDF using Eq.~\ref{eq:moments_c_3} and Eq.~\ref{eq:moment_truncated_exp}. 
From Eq.~\ref{eq:moment_truncated_exp}, we find that 
\begin{equation}
\langle c^k \rangle = \langle X^k\rangle \langle c\rangle^k.
\end{equation}
Dimensional analysis (Eq.~\ref{eq:moments_c_3}) predicts that $\langle c\rangle \sim (t/t_g)^{-3/2} =  (c_1 t/t_g)^{-3/2}$. 
The moments of concentration are therefore predicted to evolve as
\begin{equation}
 \langle c^k\rangle = (1-\xi)^k \exp\left( \frac{\xi}{1-\xi} \right) \Gamma\left(k+1,\frac{\xi}{1-\xi}\right) \left(c_1 \frac{t}{t_g} \right)^{-\frac{3}{2}k}.
 \label{eq:moment_concentration}
\end{equation}
In Fig.~\ref{fig:momentconcentration}, the predicted moments are shown with gray dashed lines, using $c_1=2.0$, which is obtained by fitting experimental and numerical data with $\xi=0.5$. The good agreement between the prediction of Eq.~\ref{eq:moment_concentration} (which assumes an exponential PDF) and the data gives further support to our proposition that the concentration PDF is exponential.
It also shows that the time evolution of all the moments of $c$, and therefore of the PDF, can indeed be accurately predicted with a single fitting parameter ($c_1$).

\begin{figure}
\centerline {\includegraphics[scale=1]{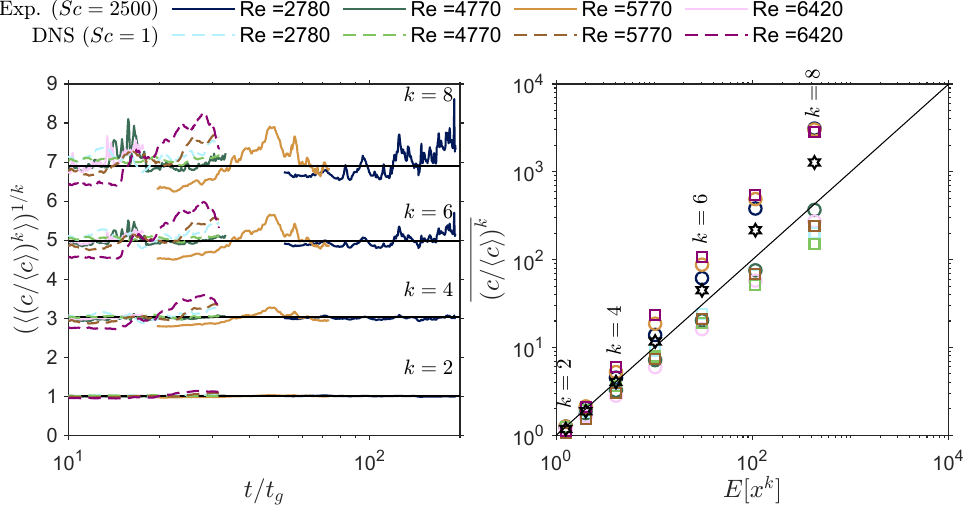}}
 \caption{(a) Moments of concentration $\langle (c/\langle c \rangle)^k\rangle^{1/k},\,\text{with}\, k=2,4,6,8$ as a function of time $t/t_g$ for the mean of three experiments at each $Re$ (solid lines) and the mean of two simulations at each $Re$ (dashed lines). The black lines show the theoretical $k$th moment $E[X^k]$ according to Eq.~\ref{eq:moment_truncated_exp}. Note that the moments are shifted along the y-axis to improve readability. (b) Temporal mean of the moment of concentration $\overline{\langle (c/\langle c \rangle)^k\rangle}$ against the predicted moment $E[X^k]$ (Eq.~\ref{eq:moment_truncated_exp}), with $k=2$ to $8$. The black stars denote the mean of the $k$th moment for all Reynolds (experiments and simulations).}
 \label{fig:moment_meanc}
\end{figure}

Fig.~\ref{fig:moment_meanc}a shows the moments of concentration $\langle (c/\langle c \rangle)^k\rangle^{1/k}$ for different $k$ as a function of time. All moments are shifted along the y-axis for better readability. The moments $\langle (c/\langle c \rangle)^k\rangle^{1/k}$ are almost constant with time in both experiments and simulations. Note that in Fig.~\ref{fig:momentconcentration}, the evolution of the moment of concentration is well predicted by Eq.~\ref{eq:moment_concentration}. 
Fig.~\ref{fig:moment_meanc}b shows the temporal mean of the moment of concentration $\overline{\langle (c/\langle c \rangle)^k\rangle}$ as a function of the predicted moment for a truncated exponential distribution (see Eq.~\ref{eq:moment_truncated_exp}). Moments of concentration are well explained by our truncated exponential distribution model, although the dispersion of the experimental and numerical values increases with $k$. Fig.~\ref{fig:moment_meanc}b shows no significant dependence on Reynolds or Schmidt numbers. This analysis applies to the finite-order moments considered here, up to $k=8$, and should not be extrapolated to the limit $k\rightarrow\infty$. In this limit, the moment is controlled by the highest concentration values. The observed increase of $c_{\max}/\langle c\rangle$ with time (Fig.~\ref{fig:meanmaxconcentration}c) indicates that rare high-concentration regions, associated with the vortex cores, do not follow the same self-similar behavior as the bulk of the concentration distribution.

\subsection{On the effect of $Sc$}
\label{subsec:Sc}

The evolution of the concentration moments in a turbulent thermal is independent of the Schmidt number $Sc$ within the range $[1,2500]$ presented here. In addition, the statistical distribution of $c$ is also, at first order, independent of the diffusivity. However, differences do exist. First, we observe thicker structures or lamellae in the simulations than in the experiments (see Figs.~\ref{fig:anatomy_exp} and ~\ref{fig:anatomy_sim}). At any time, the diffusivity length scale is 50 times larger for the simulations than for the experiments. 

Fig.~\ref{fig:PDF_meanc}b reveals that the maximum concentration (compared to the mean concentration) is smaller in the $Sc=1$ simulations compared to the $Sc=2500$ experiments. This suggests that diffusion affects the evolution of the cores as maximum values of concentration correspond to the thermal cores (see Fig.~\ref{fig:anatomy_exp}c). 

The exponential nature of the concentration PDF is unaffected by the diffusivity, as it is primarily determined by intermediate concentration values. Extreme values of concentration, which deviate from the exponential and are contained in the cores of the turbulent thermal, might be preserved more effectively in turbulent thermals with higher Péclet numbers (lower diffusivity). A more thorough study of the evolution of the cores will be required to clarify the role of the mixing processes in the long-term persistence of the cores.

\clearpage
\newpage
\section{Discussion and conclusion}
\label{sec:discussion}

Early work on turbulent thermals --- buoyant masses rising or sinking in a quiescent environment --- primarily investigated their integral evolution, focusing on the concept of turbulent entrainment \citep{b1954,mtt1956,t1986,lj2019}. More recently, experimental studies examined the internal structure of thermals by measuring their density and vorticity content \citep{bj2005,bj2010,zlla2013}. However, these studies have not examined the concentration distributions in turbulent thermals, which is an insightful tool for investigating their internal structure and mixing dynamics.

In this study, we performed laboratory experiments and numerical simulations of turbulent thermals and analyzed the evolution of the moments and distribution of concentration. Our results show that the concentration fields of turbulent thermals are self-similar in time, except at high concentrations. Far from the source, the concentration distribution, normalized by the average concentration, is independent of time in both experiments and simulations (Fig.~\ref{fig:PDFc_meanc_expsim}). This leads to normalized moments of concentration that are constant over time (Fig.~\ref{fig:moment_meanc}a). As discussed in Section~\ref{sec:introduction}, this self-similar behavior aligns with the fact that both $Re$ and $Pe$ remain constant over time when defined with the instantaneous radius and velocity (Eq. \ref{eq:AD1}), resulting in constant normalized concentration moments in the asymptotic regime (Eq. \ref{eq:moments_c_2.5}). Self-similarity also requires that the typical length scales of concentration heterogeneities remain at a fixed ratio of the thermal size, and therefore evolve in proportion to $r$. The natural length scales of the problem --- thermal size, diffusion length, and the Batchelor scale $\ell_{B}=(D/\gamma)^{1/2}$, where $\gamma$ is the stretching rate of the concentration lamellae --- all appear to have the same time dependence when far from the source, as shown below. The thermal radius increases as $t^{1/2}$ as predicted by equation \ref{eq:r_1}. This is also the case for the diffusion length $(Dt)^{1/2}$.
Dimensional analysis along the same lines as in Section \ref{subsection:BulkProperties} shows that, far from the source, the stretching rate should be of the form $\gamma = f_{\gamma}(Re)/t$, where $f_{\gamma}(Re)$ is an unknown function of $Re$. Therefore, the Batchelor length scale $\ell_{B}$ evolves as $ f_{\gamma}(Re)^{-1/2} (Dt)^{1/2}$.

The stationary nature of the distribution of $c/\langle c \rangle$, i.e., the self-similar behavior, can be qualitatively understood as resulting from a balance between the competing effects of homogenization within the thermal and continuous addition of ambient liquid through engulfment. This is well illustrated by the fact that the ratio of the standard deviation to the mean of concentration $\sqrt{\sigma^2}/\langle c \rangle$ is constant with time (equal to 1 for an exponential distribution), which contrasts with the evolution of stirred concentration fields in confined spaces, for which the ratio $\sqrt{\sigma^2}/\langle c \rangle $ tends toward $0$ as mixing proceeds. However, the details of the mechanisms by which the stationary distribution is maintained have yet to be understood and modeled.

Our interpretation of self-similarity contrasts with the findings of \citet{zlla2013}, who found that the overall thermal does not show a self-similar evolution, although the cores do evolve self-similarly. This conclusion relies on a different measure of self-similarity, where the stability of ensemble-averaged radial vorticity and density profiles is assessed over the development of the thermal. The profiles are normalized by their maximum value, located in the cores, which differs from our results, where we normalize the concentration distribution by the average concentration of the thermal. Since the maximum concentration decreases more slowly than the average concentration, this may explain why self-similarity is observed in the cores but not in the overall thermal. Normalizing the profiles by quantities averaged over the thermal could result in a self-similar evolution of the profiles, except in the core, which is consistent with our findings.

Our results show that the concentration distribution is well approximated by an exponential PDF (Fig.~\ref{fig:PDF_meanc} and \ref{fig:moment_meanc}b), except at the largest concentrations where the distribution becomes more scattered. This exponential behavior is reminiscent of similar features observed in other flow configurations, such as the exponential tails in expanding systems (turbulent jets \citep{div2010}) and in confined settings (turbulent plumes \citep{vd2003,dv2008} or porous-media flows \citep{ldv2015}). In contrast, the PDF obtained here is stationary when rescaled by the time-dependent mean --- a notable difference from these cases --- owing to the continuous engulfment of fresh ambient fluid by the thermal. The concentration distribution characterizes the fluid mixing processes, which can be interpreted in terms of the evolution of concentration lamellae \citep{v2019}, governed by their stretching histories and by interactions between lamellae (whether they remain isolated or aggregate). In several related configurations, models describing lamellar stretching and interaction have been successfully developed \citep{dv2008,ldv2013,ldv2015,lhdv2017}. Identifying the mechanisms responsible for both the exponential shape of our PDF and its self-similarity will require further investigation to establish a quantitative model for the elongation dynamics.

We also observed a departure from exponential behavior, which is related to the high-concentration cores of the thermal and suggests locally different mixing dynamics. The cores of the thermal are protected from mixing, whereas the peripheral region is mixed more efficiently (Fig.~\ref{fig:evolution_exp_sim}). This separation of dynamics --- where the core is ''protected'' while the periphery is mixed --- echoes the discussion by \cite{rotily2025}. They highlight that in regions of high coherence, such as the vortex cores of our thermals, transport is likely governed by the conservation of vorticity, which effectively sequesters the scalar. In contrast, the peripheral regions follow a more 'Prandtl-like' momentum exchange, where continuous engulfment facilitates rapid homogenization. Their discussion is consistent with our two observations: a slower decrease in the maximum concentration than in the mean concentration, and slower decrease in the experiments compared to the simulations. For our simulations with $Sc \sim 1$, concentration may "leak" outside the vortex cores due to the broader concentration profiles compared with the velocity profiles \cite{rotily2025}. In contrast, for our experiments with $Sc \gg 1$, the concentration would be "sequestered" inside the vortex cores \cite{rotily2025}.

An unexpected observation is the very weak effect of diffusion on the PDFs of concentration, as shown by the comparison of experiments and numerical simulations carried out at the same values of the Reynolds number, but differing by a factor of 2500 in terms of Péclet numbers. In both cases, the concentration distribution is well described by the same exponential PDF. Across the range of Reynolds and Péclet numbers examined in our experiments and simulations, the concentration distributions are mostly independent of $Re$ and $Pe$ (Figs.~\ref{fig:PDF_meanc} and \ref{fig:moment_meanc}), showing a weak effect of viscosity and diffusion on the distributions. This is consistent with the recent analysis by \cite{rotily2025}, who suggests that the difference between momentum ($u$) and scalar ($c$) transport is a mixing problem governed by the way eddies exchange mass with the environment. Their work demonstrates that once the flow reaches a regime where the ''smooth reservoir'' of the environment is accessible via turbulent coarsening, the specific molecular diffusivity has a vanishingly small effect on the overall transport law. At the same $Re$, the concentration maximum is higher in experiments ($Sc=2500$) than in numerical simulations ($Sc=1$), suggesting a marginal dependence on $Pe$, possibly through diffusive homogenization of the cores. However, diffusivity strongly affects the spatial structure of the concentration field, as shown by notably finer structures in the experiments than in the simulations (Fig.~\ref{fig:evolution_exp_sim}c-d). Further analyses of the distribution of the concentration increments and their moments, known as structure functions \citep{lhdv2017,v2019}, would help characterize the mixing process that underlies the exponential distribution.

\begin{acknowledgments}
LH thanks Daniel Lecoanet for helping implement Dedalus into the HPC of Université Grenoble Alpes and for the data provided for the large-resolution simulations. All the 3D simulations presented in this paper were performed using the GRICAD infrastructure (https://gricad.univ-grenoble-alpes.fr), supported by Grenoble research communities. LH gratefully acknowledges support from the Science and Technology Facilities Council for grant ST/Y001206/1. VL was partly funded by the Horizon Europe ERC Grant 101098375\textbackslash SOFT‐PLANET. This work was supported by the European Research Council (ERC) under the European Union's Horizon 2020 research and innovation program (grant number 716429). ISTerre is part of Labex OSUG@2020 (ANR10 LABX56). 
\end{acknowledgments}


\clearpage\newpage
\appendix*

\section{Intensity--Concentration Calibrations}\label{appA}
The determination of the concentration field uses a relationship between the fluorescence intensity of Rhodamine 6G and its concentration. This relationship is independently calibrated for each experimental case (A, B, C, and D), as the initial concentrations vary from one case to another, and the fluorescence intensity slightly depends on the NaCl concentration. For each case, the fluorescence intensity of the solution contained in the tube is measured under conditions identical to the experiments to account for potential absorption effects along the path of the laser sheet. The initial aqueous solution is gradually diluted with water to consider the concurrent variation in salt concentration along with the variation in fluorescent dye concentration.
\begin{figure}
\centering
\includegraphics[scale=1]{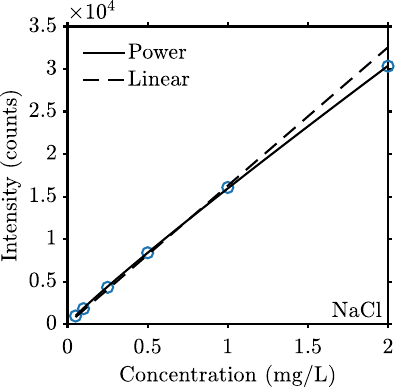}
\caption{Fluorescence light intensity as a function of Rhodamine 6G concentration in aqueous solutions of NaCl. The solid line corresponds to a power law of the form $I=ac^b$ fitted to the experimental data, with $a=(1.60\pm0.01)\times\num{1e4}$ and $b=0.93\pm0.01$. The dashed line corresponds to a power law of the form $I=ac$ fitted to the experimental data for $c \leq 1~\si{\milli\gram/\liter}$, with $a=(1.63\pm0.02)\times\num{1e4}$.}
\label{fig:calibration_concentration_01}
\end{figure}
Fig.~\ref{fig:calibration_concentration_01} shows one of the calibrations of the light intensity $I$ as a function of the concentration $c$ of Rhodamine 6G. When $c \to 0$, the light intensity should increase linearly with the dye concentration and be attenuated at high concentrations. This is indeed the case when $c \gtrsim 1~\si{\milli\gram/\liter}$. This attenuation could also be related to interactions with NaCl, leading to the formation of aggregates that reduce the fluorescence intensity. To account for this effect at high concentrations, the intensity-concentration calibration thus employs a power law of the form $I=ac^b$ rather than a linear law.

To correct for light absorption by the dye, we applied the
Beer--Lambert law along each vertical laser path. The measured and absorption-corrected intensities are related by
\begin{equation}
    I_{\mathrm{corr}}(z)
    =
    I_{\mathrm{meas}}(z)\,\exp{(A(z))},
    \qquad
    A(z)
    =
    \epsilon_{\mathrm{dye}}
    \int_{0}^{H} c_{\mathrm{before}}(z)\,\mathrm{d}z ,
\end{equation}
where $\epsilon_{\mathrm{dye}}11.6\times10^{4}$~L~mol$^{-1}$~cm$^{-1}$
is the molar absorption coefficient \citep{taniguchi2018database} and $c_{\mathrm{before}}$ is the
preliminary concentration field inferred from the measured intensity. The integration starts at the lower boundary, where the laser enters the experimental tank, and is evaluated independently along each vertical laser path. The corrected intensity is then converted back into concentration using the same calibration law, described above.

\bibliography{biblio}

\end{document}